\documentclass[12pt]{JHEP3}
\usepackage{amsfonts,epsfig}

\def\be{\begin{equation}}
\def\ee{\end{equation}}
\def\bea{\begin{eqnarray}}
\def\eea{\end{eqnarray}}
\def\bei{\begin{itemize}}
\def\eei{\end{itemize}}
\def\bee{\begin{enumerate}}
\def\eee{\end{enumerate}}
\def\noN{\nonumber}
\def\ccr{\nonumber\\}

\def\lx{\left}
\def\rx{\right}
\def\la{\langle}
\def\ra{\rangle}

\def\s{\sigma}
\def\t{\tau}

\def\mn{{\mu\nu}}

\def\ccr{\nonumber\\}

\def\s{\sigma}

\def\half{{1\over 2}}
\def\Tr{{\rm Tr}\,}

\def\non{\nonumber}
\def\fourth{{1\over4}}

\def\e{\mbox{e}}

\newcommand{\bear}{\begin{eqnarray}}
\newcommand{\ear}{\end{eqnarray}\noindent}
\newcommand{\benn}{\begin{enumerate}}
\newcommand{\enn}{\end{enumerate}}

\newcommand{\Det}{{\rm Det}}

\author{Fiorenzo~Bastianelli $^{a,b}$, Olindo~Corradini $^{b,c}$,
  Pablo~A.~G.~Pisani $^{d,e}$ and Christian Schubert $^{a,c}$\\ \vskip2mm
\hspace{-12pt} $^{a}$ Max-Planck-Institut f\"ur Gravitationsphysik, Albert-Einstein-Institut,\\
M\"uhlenberg 1, D-14476 Potsdam, Germany \vspace{3pt} \\ 
\hspace{-12pt} $^{b}$ Dipartimento  di Fisica, Universit{\`a} di Bologna
and  INFN, Sezione di Bologna,\\
Via Irnerio 46, I-40126 Bologna, Italy\vspace{3pt}\\ 
\hspace{-12pt} $^{c}$ Instituto de F\'{\i}sica y Matem\'aticas,\\
Universidad Michoacana de San Nicol\'as de Hidalgo,\\
Edificio C-3, Apdo. Postal 2-82,
C.P. 58040, Morelia, Michoac\'an, M\'exico\vspace{3pt}\\
\hspace{-12pt} $^{d}$ Theoretisch-Physikalisches Institut, Friedrich-Schiller-Universit\"at,\\
Max-Wien-Platz 1, 07743 Jena, Germany\vspace{3pt}\\
\hspace{-12pt} $^{e}$ IFLP (CONICET), Departamento de F\'isica
de la Universidad\\
Nacional de La Plata, c.c. 67, 1900 La Plata,
Argentina\\ 

E-mail:
\email{bastianelli@bo.infn.it}, \email{corradini@bo.infn.it}%\hskip.5cm  
\\ \hskip1.45cm\email{pisani@obelix.fisica.unlp.edu.ar}, \email{schubert@ifm.umich.mx}}

\abstract{The worldline formalism has in recent years emerged as a powerful tool
for the computation of effective actions and heat kernels. 
However, implementing nontrivial boundary conditions in this formalism
has turned out to be a difficult problem. Recently, 
such a generalization was developed for the case of a scalar field on the 
half-space ${\mathbb R}_+\times {\mathbb R}^{D-1}$, 
based on an extension of the associated worldline path
integral to the full ${\mathbb R}^D$ using image charges. We present here an improved
version of this formalism which allows us to write down non-recursive
master formulas for the $n$ - point
contribution to the heat kernel trace of a scalar field on the half-space with Dirichlet
or Neumann boundary conditions. These master formulas are suitable to computerization.
We demonstrate the efficiency of the formalism by a calculation of two new heat-kernel
coefficients for the half-space, $a_4$ and $a_{9/2}$.}

\preprint{AEI-2008-054, UMSNH-IFM-F-2008-25}

\title{Scalar heat kernel with boundary in the worldline formalism}

\begin{document}

\section{Introduction}
\renewcommand{\theequation}{1.\arabic{equation}}
\setcounter{equation}{0}

The ``string-inspired'' worldline formalism, originally developed in the context of
QCD scattering amplitudes \cite{berkos,strassler}, has during the last fifteen years
evolved also into a powerful tool for the calculation of effective actions
and heat kernels. One may recall that, generally, one-loop effective actions can be
written in terms of determinants of certain differentials operators depending
on given background fields. For the simplest and prototypical case, a (real) scalar field
with self-interaction $U(\phi)$, this operator is~\footnote{We use
  euclidean conventions throughout this paper.} 
\bear
H &=& -\square + U''(\phi)
\label{DefO}
\ear
and the one-loop effective action can be written as 
\bear
\Gamma [\Phi]&=& \half \ln \Det\, H
= -\half \int_0^{\infty}{dT\over T} \Tr \, \e^{-TH}
\non\\
&=&
-  \half \int_0^{\infty}{dT\over T} \int d^Dx\, K(T;x,x)
\label{Gammaphi}
\ear
where $K(T;x,x)$ is the diagonal of the heat kernel of the operator $H$,
\bear
K(T;x,y) &=& \la x | \e^{-TH} | y \ra 
\label{defK}
\ear
A standard way of calculating the effective action is given by the (diagonal) heat kernel
expansion,
\bear
K(T;x,x) &=& (4\pi T)^{-{D\over 2}}\sum_{n=0}^{\infty}a_n(x,x)\,T^n 
\label{hkexp}
\ear
See \cite{ball,vassilevich,kirsten} for reviews on the results which have been obtained
for the heat kernel coefficients $a_n$ in various field theories, as well as for
their applications in quantum field theory. 

In the worldline formalism, the starting point is the following
worldline path integral representation of the effective action (\ref{Gammaphi})
(see, e.g., \cite{review})
\bear
\Gamma[\phi] &=& - \half \int_0^{\infty}{dT\over T}\int {\cal D}x \,
\exp\biggl[-\int_0^Td\tau
\Bigl(\fourth \dot x^2 + V(x(\tau))\Bigr)\biggl]
\label{pi}
\ear 
where we now denote $U''(\phi(x)) =: V(x)$. Here the path integral in
(\ref{pi}) is over the space of 
all closed loops $x(\t)$ in spacetime with periodicity $x(T) =
x(0)$. It will be convenient to rescale 
the worldline action to the circle of unit length,
\bear
\int_0^Td\tau\,
\Bigl(\fourth \dot x^2 + V(x(\tau))\Bigr)
&=&
{1\over 4T} \int_0^1 d\t\, \dot x^2
+ T \int_0^1 d\t \, V(x(\t))
\label{rescale}
\ear 
In the application of (\ref{pi}) to the calculation of the effective action,
the most straightforward approach is to expand the interaction exponential, 
Taylor expand $V$, and evaluate the resulting Gaussian integrals
using a worldline Green's function adapted to the periodic boundary conditions. 
Since for periodic boundary 
conditions the path integral has perturbatively a zero mode, before doing so one has to split
\bear
x(\t) &=& x+y(\t) \nonumber\\[2mm]
\int {\cal D}x &=& \int d^Dx \int {\cal D}y
\label{split}
\ear
This leaves a dependence of the worldline Green's function 
on the boundary conditions
chosen for $y(\t)$. The main two choices are

\benn

\item
(Worldline) Dirichlet boundary conditions (DBC), $y(0)=y(1)=0$. This
  leads to a worldline correlator 
$\la y^{\mu}(\t)y^{\nu}(\s) \ra = -T\delta^{\mn}\,G_D(\t,\s)$, where
\bear
G_D(\t,\s) &=& 2\t(\s-1) \theta(\s-\t) + 2\s(\t-1) \theta(\t-\s)
\noN\\&=& |\t-\s|+\frac{1}{2}(1-2\t)(1-2\s)-\frac{1}{2}
\label{GD}
\ear

\item
``String-inspired'' boundary conditions (SI), $\int_0^1 d\t\, y(\t) =0$. This yields a correlator
$\la y^{\mu}(\t)y^{\nu}(\s) \ra = -T\delta^{\mn}\,G_S(\t,\s)$ with
\bear
G_S(\t,\s) &=& |\t-\s| - (\t-\s)^2 
\label{GB}
\ear
where the coincidence limit has been subtracted; see discussion in~\cite{fhss,review}.  
\enn
\noindent
After this zero mode fixing, it is natural to Taylor expand $V$ at the point $x$,
\bear
V(x+y) &=& \e^{y\cdot\partial}V(x)
\label{taylor}
\ear
Combining (\ref{taylor}) with the expansion of the interaction exponential 
renders the path integral Gaussian. Formal Gaussian integration using
either of the worldline correlators (\ref{GD}) or 
 (\ref{GB}) then leads to the following master formula
for the effective action,
\bear
\Gamma[\Phi] &=&
-\half
\int_0^{\infty}{dT\over T}
\int d^D x \, 
K_{D,S}(T,x)
\label{master}
\ear
where
\bear
K_{D,S}(T,x)
&=&
(4\pi T)^{-{D\over 2}}
\sum_{n=0}^{\infty}
{(-T)^n\over n!}
\int_0^1d\tau_1 \cdots \int_0^1d\tau_n
\nonumber\\
&& 
\hspace{-10pt}\times
{\rm exp}\biggl[-{T\over 2}
\sum_{i,j=1}^n G_{D,S}(\tau_i,\tau_j)\partial_i\cdot\partial_j
\biggr]
\, V^{(1)}(x)\cdots V^{(n)}(x)
\label{KDS}
\ear
Here it is understood that the derivative $\partial_i$ acts on $V^{(i)}$.
The prefactor $(4\pi T)^{-{D\over 2}}$ represents the free path integral determinant.
As discussed in \cite{fhss} (see also \cite{review}), using the Dirichlet Green's
function (\ref{GD}) in (\ref{master}) reproduces precisely the heat kernel diagonal,
\bear
K_D(T,x) &=& K(T;x,x)
\label{KK}
\ear
On the other hand, $K_S(T,x)$
differs from $K(T;x,x)$ by terms which are total derivatives.
As it turns out, these total derivative
terms have a simplifying effect in the sense that they lead to
a more compact form of the effective action at higher orders of the
heat kernel expansion \cite{fss,winnipeg}. 
The master formula (\ref{master}) with the string-inspired Green's function has been used in
\cite{winnipeg} for a calculation of $\Gamma[\Phi]$ to order $O(T^8)$.
Both approaches have been extended to the effective action for 
quantum electrodynamics \cite{qed}, nonabelian gauge theory \cite{fhss,gaugetheory}
and gravity \cite{gra1,gra2}. 
It must be mentioned, though,
that the issue of the zero mode fixing becomes a much more nontrivial one in the
curved space case \cite{bacozi}. 

All the work cited above pertains to the standard heat kernel, derived from operators
defined on ${\mathbb R}^D$ or on a manifold without boundary. However, many
important physics applications of effective actions involve nontrivial boundary conditions. 
The prime example is Casimir energies,
for whose calculation there is presently still no sufficiently general method
available in standard QFT, while there is increasing motivation 
not only from QED (see~\cite{Bordag:2001qi} for a review) but also from the physics of branes and
field theories with extra dimensions. In particular,
for a non supersymmetric brane configuration the Casimir force
is an important ingredient for the analysis of the stability
of the configuration~\cite{brane-casimir}. 
During the last few years a variant of the worldline formalism 
based on a direct numerical calculation of the path integral \cite{wlmontecarlo}
has been applied very successfully 
to the calculation of Casimir energies with Dirichlet boundary conditions \cite{wlcasimir}.
This strongly suggests that one should study how to implement
boundary conditions also in analytic versions of this formalism. 
Similarly, an important ingredient in the discovery of
Ho{\v r}ava-Witten duality~\cite{HW}, a cornerstone of string theory, 
was the cancellation of Einstein anomalies in the ten-dimensional boundary of
${\mathbb R}^{10}\times {\bf S}^1/{\mathbb Z}_2$. Worldline models
have so far been quite successfully used in the computation of anomalies in boundaryless 
manifolds~\cite{wlanomalies,gra1}, so
that a generalization to spaces with boundary seems in order.

For manifolds with a boundary, the expansion formula for the heat
kernel trace
\bear
\Tr \e^{-T H} =(4\pi T)^{-\frac D2} \sum_{n} a_n T^n
\ear
remains valid but now the integrated coefficients $a_n$ include
boundary contributions, and also half-integer values of $n$ appear~\cite{McKean:1967}.
Various methods for computing the heat kernel trace on manifolds with boundary
can be found in the literature 
\cite{McKean:1967,Kennedy:1979ar,Branson:1990a,Cognola:1990kq,McAvity:1990we,Branson:1995cm,
  Branson:1999jz};   
see also \cite{vassilevich,kirsten} for reviews oriented towards physical applications. 

As a first step in the direction of a generalization of worldline
techniques to spaces with boundary, in 
\cite{Bastianelli:2006hq} three of the present authors
considered the heat kernel for a scalar field propagating on the
half-space ${\mathbb R}_+\times {\mathbb R}^{D-1}$ and used the image charge method to
write the heat kernel trace on the half space as a combination of
heat kernels on the whole space. 
For the simple half-line ${\mathbb R}_+$ (that already captures the main features of the
method)  with Dirichlet/Neumann boundary conditions this yields the following
combination of whole line heat kernels  
\bea
{\rm Tr}\,\e^{-T H} = \int_0^\infty dx\, \langle x\mid \e^{-TH}\mid x\rangle
\mp \int_0^\infty dx\, \langle -x\mid \e^{-TH}\mid x\rangle
\label{halfline}
\eea   
Here the upper (lower) sign corresponds to the Dirichlet (Neumann) case.
In the following the first term will be called the ``direct'' contribution,
the second one the ``indirect'' one. In the path integral picture the second
term represents the contribution to the heat kernel due to paths which involve
a reflection at the boundary.
The calculation of these whole line heat kernels involves the
``doubled'' potential
\bea
V(x) \rightarrow {\tilde V(x)} &=& \theta(x) V(x) +
\theta(-x) V(-x) \non\\
&=&
V_+(x) + \epsilon(x) V_-(x)
\label{modpot}
\eea
where $\epsilon(x)$ is the ``sign'' function and $V_\pm$ indicates the
even/odd part of the potential. In \cite{Bastianelli:2006hq} this formalism
was used to reproduce the known heat kernel coefficients for the half-space, as
well as obtain two new ones, $a_3$ and $a_{7/2}$.

In the present paper, we present a more efficient and systematic approach along the same
lines. The main improvements over \cite{Bastianelli:2006hq} are the following:

\benn

\item
We represent the sign function appearing in the doubled potential $\tilde V$ (\ref{modpot})
through its Fourier transform,
\bea
\epsilon(x) = \int_{-\infty}^{\infty} \frac{dp}{\pi p} \sin(p x)    
\label{fourier}
\eea

\item
We use a path integral with antiperiodic boundary conditions for the calculation of the
indirect contribution.

\enn

The organization of this paper is as follows. In section \ref{indirect} 
we explain a general method for calculating the indirect contribution
to the half-line heat kernel, 
\bear
K^{\rm ind}_{\partial M}(T) &:=&
\int_0^\infty dx\ 
\la -x \mid \e^{-TH}\mid x\ra
\label{defKind}
\ear  
We use the path integral formulation to derive a master formula for this indirect part
which generalizes (\ref{KDS}). 
We elaborate this master formula for the one and two point functions,
i.e. the terms in $K^{\rm ind}_{\partial M}$ 
involving one or two $V$'s, and outline a procedure which allows one
to obtain, recursively in the number of $V$'s, a more explicit
integral representation for this indirect part.  
Section \ref{direct} presents the more intricate procedure for calculating the direct part,
\bear
K^{\rm dir}(T)&:=&
\int_0^{\infty}dx\, \la x \mid \e^{-TH}\mid x\ra
 =  \int_0^{\infty}dx\, K^{\rm dir}_{M}(T,x)+ K^{\rm dir}_{\partial
   M}(T) 
\label{defKdir}
\ear  
While the indirect part of the heat kernel contains only boundary terms, 
and will therefore be denoted by $K^{\rm ind}_{\partial M}$, the direct one 
yields both a bulk contribution $K^{\rm dir}_{M}$ and a boundary one $K^{\rm dir}_{\partial M}$. 
Again we demonstrate the efficiency of the method by an explicit treatment of the one and
two point cases. 
We summarize our results in section \ref{conclusions}.
In appendix \ref{coeffs} we
use our formalism for obtaining two more coefficients for the half-line heat kernel than were
known before, $a_4$ and $a_{9/2}$.
The generalization of the method from the half-line to a half--space is straightforward,
and is presented in appendix \ref{halfspace}. 

%                                                                 %
%% \section{Heat kernel on the half-line: Indirect contribution} %%
%                                                                 %
%\vspace{20pt}
\section{Heat kernel on the half-line: Indirect contribution}
\label{indirect}
\renewcommand{\theequation}{2.\arabic{equation}}
\setcounter{equation}{0}

Let us thus reconsider the calculation of the heat kernel for
the half-line (\ref{halfline}). We will start with the indirect term,
since, as will be seen, in the present approach it is easier to obtain than
the direct one (this was different in the approach of \cite{Bastianelli:2006hq}).

We first rewrite, using the symmetry $x \leftrightarrow -x$,
\bea
K_{\partial M}^{\rm ind}(T) 
&=&\int_0^{\infty} dx\, 
\la -x \mid \e^{-TH}\mid x\ra
= {1\over 2}  \int_{-\infty}^{\infty} dx\,  
\la -x \mid \e^{-TH}\mid x\ra
\nonumber\\
&=&
{1\over 2} \int_{ABC} {\cal D} x \,
\exp\Bigl [ -{1\over 4T}\int_0^1 d\t \dot x^2 -T \int_0^1d\t 
\, \tilde V(x(\t))\Bigr ]
\nonumber\\
&=&\fourth \,
 \Bigl\la \e^{-T \int_0^1 d\t\, \tilde
  V(x(\t))}\Bigr\ra_{ABC}  
\label{piABC}
\eea
where now the path integral is to be evaluated with {\it antiperiodic}
boundary conditions, $x(1)=-x(0)$. Here in the last equation we have
used the free antiperiodic path integral determinant which is easily
obtained by, e.g., $\zeta$ - function regularization. One finds
\bear
\int_{ABC} {\cal D} x \,\,
\e^{-{1\over 4T}\int_0^1 d\t\, \dot x^2}
&=& 
\half
\label{freedetABC}
\ear 
We also note that in the antiperiodic case
there is no zero mode, no residual integration and therefore 
also no ambiguity in the Green's function. The appropriate worldline correlator is
\bea
\la y^{\mu}(\t)y^{\nu}(\s) \ra &=& -T\delta^{\mn}\,G_A(\t,\s),\non\\[2mm]
G_A (\t,\s) &=&  |\t -\s| - {1\over 2}
\label{defDeltaA}
\eea
Note that $G_A$ is antiperiodic in
both arguments. 

We expand out the interaction exponential in (\ref{piABC}). Using (\ref{modpot}),
(\ref{fourier}), and Taylor expanding each 
$\tilde V^{(i)}(x(\tau_i))$ at
the boundary $x=0$, we can write 
\bea
\tilde V^{(i)}(x_i) &=& V^{(i)}_+(x_i) + {1\over\pi} 
\int_{-\infty}^{\infty} {dp_i\over p_i}\sin(p_ix_i) V^{(i)}_-(x_i)
\nonumber\\
&=&
{\rm e}^{x_i\partial_i}V_+^{(i)}(0)
+{1\over \pi i}\int_{\rm ev} {dp_i\over p_i}\,
\,{\rm e}^{x_i(ip_i+\partial_i)}V_-^{(i)}(0)
%\Bigr\vert_{\rm even(p_i)}
\label{splitVi}
\eea
Here we denoted $x(\tau_i)=:x_i$, and we have introduced the
abbreviation $\int_{\rm ev} dp_i$ for the integral $\int_{-\infty}^{\infty}dp_i$
with the understanding that  
only even powers of $p_i$ are to be kept in the integrand.
Applying this procedure to 
the correlator (\ref{piABC}), we obtain the following master formula
for the indirect term,
\bea
%\int_0^{\infty} dx \la -x,T|x,0\ra
K^{\rm ind}_{\partial M}
&=&
{1\over 4}
\sum_{n=0}^{\infty}{(-T)^n\over n!}
\int_0^1d\tau_1 \cdots \int_0^1d\tau_n
\,\exp \Bigl[-{T\over 2} \sum_{i,j=1}^n G_{A}(\t_i,\t_j)
D_i(p)D_j(p)
\Bigr]
\nonumber\\&&\hspace{-10pt}\times 
\Bigl[V_+^{(1)}(0)+{1\over\pi i}\int_{\rm ev} {dp_1\over p_1}
V_-^{(1)}(0)\Bigr]
\cdots
\Bigl[V_+^{(n)}(0)+{1\over\pi i}\int_{\rm ev}
 {dp_n\over p_n}V_-^{(n)}(0)\Bigr]
%\Bigl\vert_{\rm even (p_1,\ldots,p_n)}
\label{masterindirect}
\eea
Here and in the following we define 
$D_i(p):= \partial_i+ip_i$, and
it is understood that a $D_i(p)$ in the exponent acts as such on
$V_-^{(i)}$, but reduces to $\partial_i$ when acting on $V_+^{(i)}$.

Although this master formula could be used as it stands to compute individual
terms in the heat kernel expansion, 
it turns out to be
advantageous to trade the parameters $p_i$ for parameters $s_i$, 
defined in the following way: first, we note that for the terms
involving both $V_+$ and $V_-$ we can take the ordering to be 
$V_-^{(1)},\ldots,V_-^{(m)}$ and $V_+^{(m+1)},\ldots, V_+^{(n)}$.
We then use the elementary identity
\bear
{1\over 2\pi i}\int_{-\infty}^{\infty}
{dp\over p}\,\e^{-ap^2}(\e^{ibp}-\e^{-ibp})
&=&
{b\over \sqrt{\pi a}}\int_0^1ds\,\e^{-{b^2\over 4 a}s^2}
\label{ptos}
\ear
(valid for $a>0$)
recursively with $p=p_1,\ldots,p=p_m$ to eliminate all
the $p$ integrals and replace them by integrals 
$\int_0^1ds_1\cdots \int_0^1ds_m$.

We will now apply
this formalism to the case of the one and two point functions.

\subsection{The one-point function}

The term with $n=1$ in the right hand side of the master formula (\ref{masterindirect}) reads
\bear
K^{{\rm ind} (1)}_{\partial M}
&=& -{T\over 4}\int_0^1d\tau \exp\Bigl[-{T\over 2}\,G_A(\tau,\tau)
(\partial+ip)^2\Bigr]\nonumber\\
&&\times\Bigl[V_+(0)+{1\over \pi i}\int_{\rm ev}
{dp\over p}\, V_-(0)\Bigr]
%\nonumber\\&&\hspace{20pt} 
=: K^{\rm ind}_{\partial M+}+K^{\rm ind}_{\partial M-}
\label{masterindirect1}
\ear
Since $G_A(\tau,\tau)=-\half$ there is no $\tau$ dependence in the
one-point case, so that the $V_+$ term becomes trivial,
\bear
K^{\rm ind}_{\partial M+} = -\frac{T}{4}\,\e^{{T\over 4}\partial^2}V_+(0)
\label{IVplus}
\ear
In the second term, we use (\ref{ptos}) to transform it into
\bear
K^{\rm ind}_{\partial M-} &=& -\frac{T}{4}\,\,\e^{{T\over 4}\partial^2}
{1\over 2\pi i}\int_{\rm ev}{dp\over p}\,
\e^{-{T\over 4}p^2}\Bigl(\e^{iT\partial p /2}-\e^{-iT\partial p/2}\Bigr)
\, V_-(0)\nonumber\\
&=& -{T^{3\over 2}\over 4\sqrt{\pi}}
\int_0^1ds \,\e^{(1-s^2){T\over 4}\partial^2}
\partial V_-(0)
%\nonumber\\
\label{IVminus}
\ear
Using (\ref{modpot}) in reverse, and noting that even (odd)
derivatives of $V_-\ (V_+)$ vanish when evaluated at the boundary,
the two terms can be recombined, and our final result for the indirect 
part of the one-point function becomes
\bear
K^{{\rm ind}(1)}_{\partial M}(T) &=& -\frac{T}{4}
\biggl(\e^{{T\over 4}\partial^2} 
+ \sqrt{T\over\pi}\int_0^1ds \,\e^{(1-s^2){T\over 4}\partial^2}
\partial \biggr)V(0)
%\nonumber\\
\label{combineIV}
\ear
It is straightforward to expand the functional 
(\ref{combineIV}) and check that the coefficients of
the expansions match those of the literature 
(see, e.g., \cite{Bastianelli:2006hq}). 

\subsection{The two-point function}

Proceeding to the terms quadratic in $V$, the simplest one in 
(\ref{masterindirect}) is the one involving two $V_+$. It reads
\bear
K^{{\rm ind}}_{\partial M++} &=& {T^2\over 8}\int_0^1d\tau_1\int_0^1d\tau_2
\,\,\exp\biggl[{T\over 4}\Bigl(\partial_{1}^2+\partial_{2}^2
-4G_{A12}\partial_{1}\partial_{2}\Bigr)\biggr]
\nonumber\\&&\times V^{(1)}_+(0)V^{(2)}_+(0)
%\nonumber\\
\label{Ipp}
\ear
where we have abbreviated $G_{Aij}:=G_A(\tau_i,\tau_j)$. 
The terms involving both $V_+$ and $V_-$ become, using again (\ref{ptos}),
\bear
K^{{\rm ind}}_{\partial M+-} &=&  {T^2\over 8} \int_0^1d\tau_1\int_0^1d\tau_2
\,\,\exp\biggl[{T\over 4}\Bigl(\partial_{1}^2+\partial_{2}^2
-4G_{A12}\partial_{1}\partial_{2}\Bigr)\biggr]
\nonumber\\
&&\hspace{-10pt}\times
{1\over \pi i}\int_{\rm ev}{dp_1\over p_1}
\exp\biggl[-{T\over 4}p_1^2+iT\Bigl(\half\partial_{1}
-G_{A12}\partial_{2}\Bigr)p_1\biggr]
%\Bigl\vert_{{\rm even}(p_1)}
V^{(1)}_-(0)V^{(2)}_+(0)\nonumber\\[2mm]
&=&
{T^{5\over 2}\over 8\sqrt{\pi}}
 \int_0^1d\tau_1\int_0^1d\tau_2
\,\,\exp\biggl[{T\over 4}\Bigl(\partial_{1}^2+\partial_{2}^2
-4G_{A12}\partial_{1}\partial_{2}\Bigr)\biggr]
\nonumber\\&&
\hspace{-10pt}\times\Bigl(\partial_{1}-2G_{A12}\partial_{2}\Bigr)
%\nonumber\\&&\times 
\int_0^1ds_1
\,\exp\biggl[-{T\over 4}\Bigl(\partial_{1}-2G_{A12}\partial_{2}\Bigr)^2
s_1^2\biggr]\nonumber\\&&
\hspace{-10pt}\times V^{(1)}_-(0)V^{(2)}_+(0)
%\nonumber\\
\label{Ipm}
\\[2mm]
&=& K^{{\rm ind}}_{\partial M-+} \noN
\ear
For the term with two $V_-$, we apply (\ref{ptos}) first
to $p_1$ and then to $p_2$. This yields
\bear
K^{{\rm ind}}_{\partial M--} &=&
{T^2\over 8} \int_0^1d\tau_1\int_0^1d\tau_2
\,\,\exp\biggl[{T\over 4}\Bigl(\partial_{1}^2+\partial_{2}^2
-4G_{A12}\partial_{1}\partial_{2}\Bigr)\biggr]
\nonumber\\
&&\hspace{-10pt}\times
\Bigl({1\over \pi i}\Bigr)^2\int_{\rm ev}{dp_1\over p_1}
\int_{\rm ev}{dp_2\over p_2}
\exp\biggl\lbrace T\Bigl\lbrack -{1\over 4}(p_1^2+p_2^2)
+G_{A12}p_1p_2\nonumber\\
&&\hspace{-10pt}+i\Bigl(\half\partial_{1}
-G_{A12}\partial_{2}\Bigr)p_1
+i\Bigl(\half\partial_{2}
-G_{A12}\partial_{1}\Bigr)p_2
\Bigr\rbrack\biggr\rbrace
%\Bigl\vert_{{\rm even}(p_1,p_2)}
\nonumber\\&&\hspace{-10pt}\times
\, V^{(1)}_-(0)V^{(2)}_-(0)
\nonumber\\[2mm]
&=& 
-{T^2\over 8\pi}
\int_0^1d\tau_1\int_0^1d\tau_2
\,\,\exp\biggl[{T\over 4}\Bigl(\partial_{1}^2+\partial_{2}^2
-4G_{A12}\partial_{1}\partial_{2}\Bigr)\biggr]
\nonumber\\&&\hspace{-10pt}\times
\int_0^1ds_1{1\over\sqrt{1-4G_{A12}^2s_1^2}}
\,\exp\biggl[-{T\over 4}s_1^2\Bigl(\partial_{1}-2G_{A12}\partial_{2}\Bigr)^2\biggr]
\nonumber\\&&\hspace{-10pt}\times
\Biggl\lbrace
4G_{A12}\,\exp\biggl[-{T\over 4}\,{\Bigl(2(1-s_1^2)G_{A12}\partial_{1}
-(1-4G_{A12}^2s_1^2)\partial_{2}\Bigr)^2
\over 1-4G_{A12}^2s_1^2}\biggr]
\nonumber\\&&\hspace{-5pt}\quad
+ T\Bigl(\partial_{1}-2G_{A12}\partial_{2}\Bigr) 
\Bigl(2(1-s_1^2)G_{A12}\partial_{1}
-(1-4G_{A12}^2s_1^2)\partial_{2}\Bigr)
\nonumber\\&&\hspace{-5pt}\quad \times\int_0^1ds_2
\,\exp\biggl[-s_2^2{ T\over 4}\,{\Bigl(2(1-s_1^2)G_{A12}\partial_{1}
-(1-4G_{A12}^2s_1^2)\partial_{2}\Bigr)^2
\over 1-4G_{A12}^2s_1^2}\biggr]
\Biggr\rbrace\,
\nonumber\\&&\hspace{-10pt}\times
V^{(1)}_-(0)V^{(2)}_-(0)\nonumber\\
\label{Imm}
\ear
It is now easy to obtain from (\ref{Ipp}), (\ref{Ipm}), (\ref{Imm}) 
any desired term in the derivative expansion of the two-point function.
We give two examples. First, let us consider $K^{{\rm ind}}_{\partial M++}$. Here the 
$\tau_{1,2}$ integrals can be done, for example, by
expanding the exponential factor involving $G_{A12}$, and 
using 
\bear
\int_0^1d\tau_1\int_0^1d\tau_2\,
G_{A12}^n &=& {1\over 2^n(n+1)}
\label{intGn}
\ear
for $n$ even (remembering that odd powers of derivatives on $V_+(0)$ vanish).
Resummation yields
\bear
K^{{\rm ind}}_{\partial M++} &=& {T\over 4}\,\e^{{T\over 4}(\partial_{1}^2+\partial_{2}^2)}
\,{\sinh\Bigl({T\over 2}\partial_{1}\partial_{2}\Bigr)\over 
\partial_{1}\partial_{2}}\,\,V_+^{(1)}V_+^{(2)}
\label{Ippfin}
\ear
Second, let us extract the leading contribution (as a power in $T$)
from (\ref{Imm}); such a contribution is proportional to
$(\partial V(0))^2$. 
It is easy to see that the first term in braces in (\ref{Imm}) 
does not contribute. The second one does, and yields
\bea
&&
{T^3\over 8\pi}
\Bigl(\partial V(0)\Bigr)^2
\int_0^1 d\t_1\int_0^1 d\t_2\int_0^1 ds_1\ 
\frac{1-4s_1^2G_{A12}^2+4(1-s_1^2)G_{A12}^2}{\sqrt{1-4s_1^2 G_{A12}^2}}
\nonumber\\
&&={T^3\over 8\pi}
\Bigl(\partial V(0)\Bigr)^2
\int_0^1 d\t_1\int_0^1 d\t_2\ 
\biggl(\sqrt{1-4G_{A12}^2}+2G_{A12}\arcsin(2G_{A12})\biggr)\noN\\
&&= T^3{\frac{3}{64}}\  
\Bigl(\partial V(0)\Bigr)^2
\eea
This is one of the coefficients that in \cite{Bastianelli:2006hq} needed to be
fixed by means of a toy model.

%                                                     %
%% Heat kernel on the half-line: Direct contribution %%
%                                                     %
%\vspace{20pt}
\section{Heat kernel on the half-line: Direct contribution}
\label{direct}
\renewcommand{\theequation}{3.\arabic{equation}}
\setcounter{equation}{0}

We proceed to the more involved
calculation of the first (``direct'') term of the
heat kernel for the half-line (\ref{halfline}).
As in (\ref{piABC}), we use the
symmetry $x \leftrightarrow -x$ to make the $x$ integral run over the whole line,
\bear
K^{\rm dir}(T) &=& \int_0^{\infty} dx \, K^{\rm dir}(T,x) = 
{1\over 2}  \int_{-\infty}^{\infty} dx \, K^{\rm dir}(T,x)
\nonumber\\
&=&
{1\over 2} \int_{PBC} {\cal D} x \,
\exp\Bigl\lbrack -{1\over 4T}\int_0^1 d\t\, \dot x^2 -T \int_0^1d\t \,\tilde V(x(\t))\Bigr\rbrack
\label{piPBC}
\ear
This integral is formally identical to the one without a boundary,
eq. (\ref{pi}). Thus its calculation proceeds as in the whole line case,
leading to the standard master formula (\ref{KDS}) with $V(x)$
replaced by $\tilde V(x)$:
\bear
K^{\rm dir}(T,x)
&=&
(4\pi T)^{-{1\over 2}}
\sum_{n=0}^{\infty}
{(-T)^n\over n!}
\int_0^1d\tau_1 \cdots \int_0^1d\tau_n
\nonumber\\
&& 
\hspace{-10pt}\times
{\rm exp}\biggl[-{T\over 2}
\sum_{i,j=1}^n G_{D,S}(\tau_i,\tau_j)\partial_i\partial_j
\biggr]
\prod_{i=1}^n
\, \Bigl\lbrack V^{(i)}_+ (x) + (\epsilon (x)V_-(x))^{(i)} \Bigr\rbrack
\label{Kdirmaster}
\ear
In the case of an even potential $\tilde V(x) = V_+(x)$ the further evaluation of this
master formula would then also proceed as in the whole line case.
Things get much more involved in the presence of $V_-(x)$, since the derivatives in
(\ref{Kdirmaster}) can also act on the $\epsilon(x)$ contained in $\tilde V(x)$, and produce
$\delta$ functions and derivatives thereof. However such terms are boundary terms,
therefore the complete bulk part of the heat kernel gets produced by
the terms where all derivatives hit~$V$'s:
\bear
K_{M}^{\rm dir}(T,x)
&=&
(4\pi T)^{-{1\over 2}}
\sum_{n=0}^{\infty}
{(-T)^n\over n!}
\int_0^1d\tau_1 \cdots \int_0^1d\tau_n
\nonumber\\
&& 
\hspace{-10pt}\times
{\rm exp}\biggl[-{T\over 2}
\sum_{i,j=1}^n G_{D,S}(\tau_i,\tau_j)\partial_i\partial_j
\biggr]
\prod_{i=1}^n
\, \Bigl\lbrack V^{(i)}_+ (x) + \epsilon (x)V^{(i)}_-(x) \Bigr\rbrack
\label{Kdirbulkmaster}
\ear
Here our notation is meant to convey that 
the derivative $\partial_i$ acts only on $V^{(i)}$, not on $\epsilon (x)$.
In this bulk part one is now free to take $x$ positive and to replace 
$V^{(i)}_+ (x) + \epsilon (x)V^{(i)}_-(x)$ by $V^{(i)}(x)$.

For the explicit evaluation of the boundary part, 
we use again the Fourier representation of
the $\epsilon$ function (\ref{fourier}). This allows us to rewrite
(\ref{Kdirmaster}) as
\bear
K^{\rm dir}(T,x)
&=&
(4\pi T)^{-{1\over 2}}
\sum_{n=0}^{\infty}
{(-T)^n\over n!}
\int_0^1d\tau_1 \cdots \int_0^1d\tau_n
\nonumber\\ && 
\hspace{-10pt}\times
{\rm exp}\biggl[-{T\over 2}
\sum_{i,j=1}^n G_{D,S}(\tau_i,\tau_j)D_i(p)D_j(p)
\biggr]
\nonumber\\ && 
\hspace{-10pt}\times
\prod_{k=1}^n
\, \Bigl\lbrack V^{(k)}_+ (x) + {1\over \pi i}\int_{\rm ev}{dp_k\over p_k}
\,\e^{ip_kx}V_-^{(k)}(x) \Bigr\rbrack
%\nonumber\\
\label{Kdirmastermod}
\ear
Now (\ref{Kdirmastermod}) and (\ref{Kdirbulkmaster}) differ only in the exponential prefactor.
Rewriting this difference as the integral of a total derivative in a new variable $w$, and
integrating over $x$, we
obtain the following master formula for 
$K_{\partial M}^{\rm dir}$:
\bear
K_{\partial M}^{\rm dir}(T)
&=& 
\half \int_{-\infty}^{\infty} dx\, \Bigl(K^{\rm dir}(T,x)-K_M^{\rm dir}(T,x)\Bigr)
\non\\
&=& 
\half
(4\pi T)^{-{1\over 2}}
\sum_{n=0}^{\infty}
{(-T)^n\over n!}
\int_0^1d\tau_1 \cdots \int_0^1d\tau_n
\int_{-\infty}^{\infty}dx
\nonumber\\
&& \hspace{-10pt}
\times \int_0^1dw \, \frac{\partial}{\partial w}
\,{\rm exp}\biggl[-{T\over 2}
\sum_{i,j=1}^n G_{D,S}(\tau_i,\tau_j)D_i(wp)D_j(wp)
\biggr]
\non\\
&&\hspace{-10pt}\times
\prod_{k=1}^n
\, \Bigl\lbrack V^{(k)}_+ (x) + {1\over \pi i}\int_{\rm ev}{dp_k\over p_k}
\,\e^{ip_kx}V_-^{(k)}(x) \Bigr\rbrack
\label{Kdirboundarymaster}
\ear
Since this difference contains only boundary terms, it is justified to
Taylor expand $V_\pm (x)$ with respect to the boundary, $V_\pm (x)=\e^{x\partial}V_\pm (0)$. 
The $x$ integral can then
be done, and yields a $\delta$ function involving the various $p_k$'s and $\partial_k$'s. 
The $w$-derivative
always cancels one of the (spurious) poles in the $p_k$'s, after which one can use the 
$\delta$ function to eliminate the corresponding $p_k$ integral. After this one first does
the remaining $p_k$ integrals, then the $w$ integral, and finally the $\tau_i$ integrals.
The latter ones will be of Selberg type.
Here it should also be mentioned that, due to the fact that the string-inspired Green's function
preserves the worldline translation invariance, in the SI scheme it is always
possible to eliminate one of the $\tau$ - integrals trivially, i.e. by setting $\tau_n=0$.
This is generally not the case for the DBC scheme.

As for the indirect contribution, we will demonstrate the procedure by
an explicit calculation of the one and two point contributions.

\subsection{The one-point function}

At the one-point level, the bulk master formula (\ref{Kdirbulkmaster}) 
gives
\bear
K_{M}^{{\rm dir}(1)}(T,x)
&=&-T
(4\pi T)^{-{1\over 2}}
\int_0^1d\tau \,
\e^{-{T\over 2}G(\tau,\tau)\partial^2}
V(x)
\label{Kdir1bulk}
\ear
where $G(\t,\s)$ stands for either Green's function.
For the boundary term
we get from (\ref{Kdirboundarymaster}), after expanding $V_-(x)$ at $x=0$,
\bear
K_{\partial M}^{{\rm dir}(1)}(T) &=& 
-{T\over 2}
(4\pi T)^{-{1\over 2}}
\int_0^1d\tau
\int_{\rm ev}{dp\over i\pi p}\int_{-\infty}^{\infty}dx\,\e^{(ip+\partial)x} \non\\
&&\hspace{-10pt}\times
\int_0^1dw\, \partial_w \,\e^{-{T\over 2}G(\tau,\tau)(\partial+iwp)^2}
V_-(0) 
\nonumber\\
&=& 
-{T\over 2}
(4\pi T)^{-{1\over 2}}
\int_0^1d\tau
\int_{\rm ev}{dp\over i\pi p}2\pi \delta(p-i\partial) \non\\
&&\hspace{-10pt}\times
(-TG(\tau,\tau))ip\int_0^1dw\, (\partial+iwp)
\,\e^{-{T\over 2}G(\tau,\tau)(\partial+iwp)^2}
V_-(0) 
\nonumber\\
&=& 
T^2
(4\pi T)^{-{1\over 2}}
 \int_0^1d\tau\, G(\tau,\tau)
\int_0^1dw\, (1-w)\non\\&&
\hspace{-10pt}\times\, \e^{-{T\over 2}G(\tau,\tau)(1-w)^2\partial^2}\partial V_-(0)
\non\\
&=& T
(4\pi T)^{-{1\over 2}}
 \int_0^1d\tau\, \partial^{-1} \left(1-e^{-\frac T2 G(\t,\t)
   \partial^2}\right) V_-(0)
\label{Kdir1bound}
\ear
The essential point to be noted is that the total derivative in $w$ has led to the
appearance of a factor of $p$ in the numerator which cancels the one in the denominator,
making it possible to apply the $\delta$ function coming from the $x$ integral. 

Note also that, except the leading order, all terms in the expansion of both the boundary part
and the bulk part of the one-point function are
dependent upon the choice of worldline Green's function. However, let us verify
that the complete (integrated) one-point function is
scheme-independent. Combining (\ref{Kdir1bulk}) and (\ref{Kdir1bound}) we can write this as
\bea
K^{{\rm dir}(1)}(T) &=& -T(4\pi T)^{-1/2}\int_0^1 d\t\Biggl[
\int_0^\infty dx\ e^{-\frac T2 G(\t,\t) \partial^2} V(x)\nonumber\\
&&\hspace{-10pt}
+\partial^{-1}\left(e^{-\frac T2 G(\t,\t) \partial^2}-1\right)V(0) \Biggr]
\eea
Using partial integration it is thus easy to see that the latter
reduces to
\bea
K^{{\rm dir}(1)}(T) = -T(4\pi T)^{-1/2} \int_0^\infty dx\ V(x)
\eea
which is manifestly independent of the choice of Green's function.

\subsection{The two-point function}

At two points, the bulk master formula (\ref{Kdirbulkmaster}) yields 
\bear
K_{M}^{{\rm dir}(2)}(T,x)
&=& 
(4\pi T)^{-{1\over 2}}
{T^2\over 2!}
\int_0^1d\tau_1 \int_0^1d\tau_2 \,
\non\\&&\hspace{-10pt}\times{\rm exp}\Bigl[-{T\over 2}
\sum_{i,j=1}^2 G_{ij}\partial_i\partial_j
\Bigr]V^{(1)}(x)V^{(2)}(x)
%\nonumber\\
\label{Kdir2bulk}
\ear
The boundary master formula (\ref{Kdirboundarymaster}) now gives three contributions,
\bear
K_{\partial M}^{{\rm dir}(2)}
&=& 
K_{\partial M\, +-}^{{\rm dir}}
+K_{\partial M\, -+}^{{\rm dir}}
+K_{\partial M\, --}^{{\rm dir}}
\label{decompKdirbound}
\ear
where
$K_{\partial M\, +-}^{{\rm dir}}= K_{\partial M\, -+}^{{\rm dir}}$.
The mixed term is the simpler one to calculate:
\bear
K_{\partial M\, -+}^{{\rm dir}} &=& 
  {T^2\over 4}(4\pi T)^{-1/2} \int_{-\infty}^{\infty}dx \int_0^1d\tau_1\int_0^1d\tau_2 
 \,\e^{-{T\over 2}G_{22}\partial_2^2}\, \non\\
 &&\hspace{-10pt}\times
 \Bigl[
 \e^{-{T\over 2}\bigl(G_{11}(\partial_1+ip_1)^2+2G_{12}(\partial_1+ip_1)\partial_2\bigr)}
-  \e^{-{T\over 2}\bigl(G_{11}\partial_1^2+2G_{12}\partial_1\partial_2\bigr)}
\Bigr]
 \nonumber\\
 &&\hspace{-10pt}\times 
 \int_{\rm ev} {dp_1\over i\pi p_1} \e^{ip_1x}\
  V_-^{(1)}(x)V_+^{(2)}(x) \non\\
 \non\\
 &=& 
  {T^2\over 4}(4\pi T)^{-1/2} \int_0^1d\tau_1\int_0^1d\tau_2 
 \,\e^{-{T\over 2}G_{22}\partial_2^2}\, \non\\
 &&\hspace{-10pt}\times
 \int_0^1dw \partial_w 
\, \e^{-{T\over 2}\bigl(G_{11}(\partial_1+iwp_1)^2+2G_{12}(\partial_1+iwp_1)\partial_2\bigr)}
 \non\\&&
\hspace{-10pt} \times
 \int_{\rm ev} {dp_1\over i\pi p_1} \int_{-\infty}^{\infty}dx\,
 \e^{(ip_1+\partial_1+\partial_2)x}\ 
  V_-^{(1)}(0)V_+^{(2)}(0) \non\\
 \non\\
 &=& 
  {T^2\over 4}(4\pi T)^{-1/2} \int_0^1d\tau_1\int_0^1d\tau_2 
 \,\e^{-{T\over 2}G_{22}\partial_2^2}\, \non\\
 &&\hspace{-10pt}\times
 \int_0^1dw (-iTp_1)(G_{11}(\partial_1+iwp_1)+G_{12}\partial_2)
 \non\\
 &&\hspace{-10pt}\times\,
 \e^{-{T\over 2}\bigl(G_{11}(\partial_1+iwp_1)^2+2G_{12}(\partial_1+iwp_1)\partial_2\bigr)}
 \non\\&&
\hspace{-10pt} \times
 \int_{\rm ev} {dp_1\over i\pi p_1} 2\pi \delta(p_1-i\partial_1-i\partial_2)\
  V_-^{(1)}(0)V_+^{(2)}(0) \non\\
 \non\\
 &=& 
 - {T^3\over 2}(4\pi T)^{-1/2} \int_0^1d\tau_1\int_0^1d\tau_2 
 \,\e^{-{T\over 2}G_{22}\partial_2^2}\, \non\\
 &&\hspace{-10pt}\times
 \int_0^1dw \Bigl(G_{11}\bigl[(1-w)\partial_1-w\partial_2\bigr] + G_{12}\partial_2\Bigr)
 \non\\
 &&\hspace{-10pt}\times\,
 \e^{-{T\over 2}\bigl(G_{11}\bigl[(1-w)\partial_1-w\partial_2\bigr]
   ^2+2G_{12}\bigl[(1-w)\partial_1-w\partial_2\bigr]
   \partial_2\bigr)}\ V_-^{(1)}(0)V_+^{(2)}(0) 
\label{Kdir2bound-+}
\ear
The case of $K_{\partial M\, --}^{{\rm dir}}$ is somewhat more laborious:
\bear
K_{\partial M\, --}^{{\rm dir}}
&=& 
  {T^2\over 4}(4\pi T)^{-1/2}  \int_0^1d\tau_1\int_0^1d\tau_2 
  \non\\&&\hspace{-10pt}\times
\int_0^1dw \partial_w \,\e^{-{T\over 2}\sum_{i,j=1}^2G_{ij}(\partial_i + iwp_i)(\partial_j+iwp_j)}
\non\\
&&\hspace{-10pt}\times 
\int_{\rm ev} {dp_1\over i\pi p_1} 
\int_{\rm ev} {dp_2\over i\pi p_2} 2\pi\delta(p_1+p_2-i\partial_1-i\partial_2)
\ V_-^{(1)}(0)V_-^{(2)}(0) 
\non\\
&=& 
(-iT)  {T^2\over 4}(4\pi T)^{-1/2}  \int_0^1d\tau_1\int_0^1d\tau_2\int_0^1dw
  \non\\&&\hspace{-10pt}\times 
\Big\lbrack G_{11}p_1(\partial_1+iwp_1) + G_{22}p_2(\partial_2+iwp_2)
\non\\&&+G_{12}(p_1\partial_2+p_2\partial_1+2iwp_1p_2)\Big\rbrack
%\non\\&&\hspace{-10pt}\times \,
\e^{-{T\over 2}\sum_{i,j=1}^2G_{ij}(\partial_i + iwp_i)(\partial_j+iwp_j)}
\non\\
&&\hspace{-10pt}\times 
\int_{\rm ev} {dp_1\over i\pi p_1} 
\int_{\rm ev} {dp_2\over i\pi p_2} 2\pi\delta(p_1+p_2-i\partial_1-i\partial_2)
\ V_-^{(1)}(0)V_-^{(2)}(0) 
\label{Kdir2bound--}
\ear
Using the symmetry $1\leftrightarrow 2$, the expression in square brackets can be simplified to
\bear
2p_2\Big\lbrack  G_{22}\partial_2+G_{22}iwp_2+G_{12}\partial_1+iG_{12}wp_1\Big\rbrack
\label{simp}
\ear
so that the $p_2$ pole cancels and we may use the $\delta$ function 
to eliminate $p_2$ altogether. This leads to
\bear
K_{\partial M\, --}^{{\rm dir}}&=& 
-T^3(4\pi T)^{-1/2}  \int_0^1d\tau_1\int_0^1d\tau_2 \int_0^1dw\int_{\rm ev} {dp_1\over i\pi p_1} 
\non\\&&\hspace{-10pt}\times
\Bigl\lbrack
G_{22}\Bigl(\partial_2-w(ip_1+\partial_1+\partial_2)\Bigr)
+G_{12}(\partial_1+iwp_1)
\Bigr\rbrack
\non\\&&\hspace{-10pt}\times\exp\biggl\lbrace-{T\over 2}\Bigl\lbrack
G_{11}(\partial_1+iwp_1)^2 +G_{22}\Bigl(\partial_2-w(ip_1+\partial_1+\partial_2)\Bigr)^2
\non\\&&
\qquad\quad+2G_{12}(\partial_1+iwp_1)\Bigl(\partial_2-w(ip_1+\partial_1+\partial_2)\Bigr)
%\non\\&&\qquad\qquad\quad
\Bigr\rbrack
\biggr\rbrace
\non\\&&\hspace{-10pt}\times
V_-^{(1)}(0)V_-^{(2)}(0)
\label{Kdir2bound--fin}
\ear
Individual terms in the derivative expansion of this two-point function are now
obtained by expanding out the exponential factors in 
eqs. (\ref{Kdir2bulk}), (\ref{Kdir2bound-+}), (\ref{Kdir2bound--fin}).
Note that in the case of (\ref{Kdir2bound-+}) this immediately leads to a polynomial $w$ integral.
For (\ref{Kdir2bound--fin}) we have to remember the fact that
$\partial^nV_-(0) = 0$ for even $n$; it is then easily seen that all surviving terms in this
expansion carry a factor of $p_1$. The $p_1$ integral then becomes a simple Gaussian
one, and doing it one gets a $w$ integral which is polynomial. It is only for the final
$\tau_{1,2}$ integrals that one has to specify the Green's function $G(\tau_1,\tau_2)$. 
As we have seen already in the one-point case, the coefficients of a given term
will generally depend on the choice of the worldline Green's function; the equivalence
of the results obtained with different worldline Greens' functions can only be
seen after adding up bulk and boundary terms, and performing certain integration
by parts. 
However, since bulk terms generally have an even number of derivatives, the boundary terms
involved in integration by parts have an odd one. It is therefore clear that 
those terms in $K^{\rm dir}_{\partial M}$  with an even number of derivatives are 
always independent of the choice of the worldline Green's function. 

Let us exemplify all this by extracting the lowest order term in the
two-point function, i.e. the coefficient of $\partial V_-(0)\partial
V_-(0)$. In the indirect sector only
~(\ref{Kdir2bound--fin}) contributes to it.
Collecting from (\ref{Kdir2bound--fin}) all the terms involving $\partial_1
V_-^{(1)}(0)\partial_2 V_-^{(2)}(0)$, performing the
integral over $p_1$, which is now Gaussian, and
integrating out the auxiliary variable $w$ one obtains 
\bea
T^{7/2}(4\pi T)^{-1/2}\Big(\partial V(0)\Big)^2
\ \frac{1}{3\sqrt{2\pi}}\int_0^1d\t_1\int_0^1d\t_2
\Big(2G_{12}-G_{11}-G_{22}\Big)^{3/2} 
\label{dv2}
\eea 
Written in this form it is easy to see that the integrand does not depend on the
choice of worldline Green's functions~\footnote{Worldline Green's
  functions are usually defined using a linear constraint $\int_0^1 d\t \rho(\t)
  y(\t)=0$, with $\int_0^1 d\t \rho(\t)
  y(\t)=1$ (see \cite{fhss,review}). 
This definition includes as special cases the ``string inspired''
  method (with nonvanishing coincidence limit) identified by $\rho =1$, for which
  $\tilde G_S(\t-\s) = |\t-\s|-(\t-\s)^2-\frac16$, and the DBC one
  identified by $\rho(\t) =\delta(\t)$. The Green's function corresponding to a generic $\rho$
is thus related to the string inspired Green's function given in~(\ref{GB}) by the relation
$G(\t,\s) = G_S(\t-\s) -\int_0^1d\lambda\, G_S(\t-\lambda) \rho(\lambda)
-\int_0^1d\lambda\, G_S(\s-\lambda)
\rho(\lambda)+\int_0^1d\lambda\int_0^1d\mu\, 
G_S(\lambda-\mu) \rho(\lambda)\rho(\mu)$. 
Using this relation it is
simple to show the scheme-independence of~(\ref{dv2}). Also, it is
clear that the coincidence limit can be dropped from the string
inspired propagator; in general, dropping such a limit just amounts to dropping
total derivative terms from the heat kernel coefficients.}. Translation 
invariance of the SI propagator allows
one to set $\t_2=0$, leaving a single integral
\[
\int_0^1d\t_1\int_0^1d\t_2
\Big(2G_{12}-G_{11}-G_{22}\Big)^{3/2} = 2^{3/2} \int_0^1 d\t
\Big[\t(1-\t)\Big]^{3/2} = \frac{3\sqrt{2}\pi}{2^6}
\]
The indirect contribution of this term to $a_{7/2}$ becomes
\bea
\frac{T^{7/2}}{(4\pi T)^{1/2}}\frac{\sqrt{\pi}}{2}{\frac1{2^5}} \Big(\partial V(0)\Big)^2
\eea
that is in accordance with \cite{Bastianelli:2006hq}.

%\vspace{20pt}
\section{Conclusions}
\label{conclusions}
\renewcommand{\theequation}{4.\arabic{equation}}
\setcounter{equation}{0}

This work should be seen as a further step towards making the worldline formalism
useful for analytic calculations of effective actions in the presence of nontrivial
boundary conditions.  The case of a scalar field on a half-space with
Dirichlet or Neumann conditions is the natural starting point for such a program.
While in \cite{Bastianelli:2006hq} the problem of applying the worldline
formalism to this case had been solved in principle, the solution 
was not yet optimized from a technical point of view. The various improvements
which we have implemented in the present work lead to a much simpler
algorithm, which we consider very promising for future generalizations
of the analytic worldline approach to the calculation of effective actions involving
more general fields, as well as more general boundaries and types of boundary 
conditions. For example, in the case of general boundaries in curved spaces,
although it is reasonable to think that the image charges could
increase in number and even become a continuous distribution, it is also
plausible that, for points sufficiently close to a smooth boundary and
in suitable coordinates, the main contributions arise from a single
image charge. On the other hand for specific geometries, for example a suitable
curved boundary in flat space, one might find a neat set up of
image charges that could solve the problem exactly.

For the case of the half-line, 
prior to the work of \cite{Bastianelli:2006hq}
the heat kernel for an arbitrary potential $V(x)$ 
was known only up to $a_{5/2}$, to our knowledge. 
In \cite{Bastianelli:2006hq} the coefficients $a_3$, $a_{7/2}$
were obtained, although this required already
a substantial computational effort. 
As we show in appendix \ref{coeffs}, our improved formalism makes it
relatively easy to push this calculation to $a_{9/2}$. 
Moreover, the algorithms developed
in section \ref{indirect} for the indirect and in section \ref{direct}
for the direct contributions 
to the heat kernel with Dirichlet or Neumann boundary conditions 
can be easily computerized. This should allow one to obtain many
more coefficients beyond the known ones. 
In addition, one may also include more general types of boundary conditions
in the present formalism~\cite{Bastianelli:2007jr}.

It must be emphasized that the
results for the {\it local} heat kernel, i.e. the heat kernel diagonal $K(T;x,x)$, 
generally depend on
the worldline Green's function used in the calculation of the direct part. Only
when used with the `DBC' Green's function $G_D$ our algorithm yields the standard
heat kernel. The use of the `string-inspired' or other worldline Green's functions will
yield a result which is different locally, and agrees with the standard heat kernel one
only after integration over the bulk, and summation of bulk and boundary contributions,
\[
\int_0^{\infty}dx\, K_M(T,x) + K_{\partial M}(T)
\]

On the other hand, the calculation of $K^{\rm dir}$ becomes simpler
with the string-inspired Green's function,  which is therefore preferable for
effective action calculations where usually only the total effective action
is relevant.

\acknowledgments{F.B., P.P. and C.S. thank the Albert-Einstein
  Institute, Potsdam, for hospitality. 
O.C. is grateful to the Instituto de F\'isica y Matem\'aticas of the UMSNH at Morelia
for hospitality and partial support. P.P. and C.S. are grateful to INFN
and Dipartimento di Fisica of the Universit\`a di Bologna for
hospitality. The work of F.B. and O.C. was
partly supported by the Italian MIUR-PRIN contract 20075ATT78. 
 The work of P.P. was partly supported by PIP 6160, UNLP proj. 11/X381 and DAAD.}
%\newpage

\begin{appendix}

\section{Half-line heat kernel coefficients $a_4$ and $a_{9/2}$ }
\label{coeffs}
\renewcommand{\theequation}{A.\arabic{equation}}
\setcounter{equation}{0}

The indirect contributions to the $a_4$ and $a_{9/2}$ coefficients can
be obtained with the method described in section~(\ref{indirect})
and read 
\bea
a_4^{\rm ind} &=&-\frac14 V^2\partial V(0) +\frac16 \partial V \partial^2
V(0) +\frac1{12} V \partial^3 V(0)-\frac1{5!}\partial^5 V(0)\\[2mm]
a_{9/2}^{\rm ind} &=&\frac{\sqrt{\pi}}2\Biggl[ -\frac1{3!\ 2^6}\partial^6
  V(0) +\frac1{4!} V^4(0) -\frac3{2^4} \lx(\partial V\rx)^2
  V(0)-\frac1{2^3} V^2\partial^2 V(0)\nonumber\\&& +\frac1{2^5} V\partial^4
  V(0)
+\frac{5}{3\cdot 2^5}\lx(\partial^2
  V(0)\rx)^2+\frac5{2^6}\partial^3V\partial V(0)\Biggr]            
\eea
The direct contributions to the same coefficients are
obtained with the method of section~(\ref{direct}) and their boundary
contributions read
\bea
&& a_4^{\rm dir} \big|_{\partial M} =\alpha_1\, V^2\partial V(0) +\alpha_2\, \partial V \partial^2
V(0) +\alpha_3\, V \partial^3 V(0) +\alpha_4\, \partial^5 V(0)\\[2mm]
&& a_{9/2}^{\rm dir}=\frac{\sqrt{\pi}}2\Biggl[ \beta_1 \lx(\partial V\rx)^2
  V(0)+\beta_2\, \partial^3V\partial V(0)\Biggr]
\eea 
where
\bea
\alpha_1 &=&\frac14\int_0^1 d\t_1 G_{11}
%\\[2mm]&=& 
=\left\{
\begin{array}{ll}
\displaystyle{0}\ & (SI)\\[2mm]
\displaystyle{-\frac1{2^2\cdot 3 }}\ & (DBC)
\end{array}
\right.
\non\\[3mm]
\alpha_2 
&=&\frac12 \int_0^1 d\t_1\int_0^1 d\t_2\biggl[\frac12 G_{11}
  G_{22}+\frac14 G_{11}^2+G_{12}^2-
  G_{11}G_{12}\biggr]
%\nonumber\\[2mm]&=& 
=\left\{
\begin{array}{ll}
\displaystyle{\frac{2}{5!}}\ & (SI)\\[4mm]
\displaystyle{\frac4{5!}}\ & (DBC)
\end{array}
\right.\non\\[3mm]
\alpha_3 &=&\frac18\int_0^1 d\t_1 G_{11}^2
%\\[2mm]&=& 
=\left\{
\begin{array}{ll}
\displaystyle{0}\ & (SI)\\[2mm]
\displaystyle{\frac2{5!}}\ & (DBC)
\end{array}
\right.\non\\[3mm]
\alpha_4 &=&\frac1{3!\ 2^3}\int_0^1 d\t_1 G_{11}^3
%\\[2mm]&=& 
=\left\{
\begin{array}{ll}
\displaystyle{0}\ & (SI)\\[2mm]
\displaystyle{-\frac{3!}{7!}}\ & (DBC)
\end{array}
\right.\non
\eea
and
\bea
\beta_1 &=& 
-\frac{\sqrt{2}}{3\pi}\int_0^1d\t_1\int_0^1d\t_2\Big(2G_{12}-G_{11}-G_{22}\Big)^{3/2}
=-\frac1{2^5}\non\\[3mm]
\beta_2 &=& 
\frac{2\sqrt{2}}{15\pi}\int_0^1d\t_1\int_0^1d\t_2\Big(2G_{12}-G_{11}-G_{22}\Big)^{5/2}
=\frac1{2^6\cdot 3}
\non
\eea
that are scheme-independent. We have checked that the complete $a_4$
coefficient is independent of the propagator chosen (SI or DBC), as expected.

%                                 %
%% Heat kernel on the half-space %% 
%                                 % 
\section{Heat kernel on the half-space}
\label{halfspace}
\renewcommand{\theequation}{B.\arabic{equation}}
\setcounter{equation}{0}

Let us consider the flat space 
${\cal M}=  {\mathbb R}_+ \times {\mathbb R}^{D-1}$
with coordinates $X^\beta :=(x^0,x^b):=(x^0,\vec{x}) $, where 
$0 \leq x^0<\infty$ and $\vec{x} \in {\mathbb R}^{D-1}$.
The generalization of our algorithm from the half-line to the half-space
is straightforward, starting with the generalization of~$\tilde V$,
\bear
\tilde V(X) &:=& \theta(x^0)V(x^0,\vec x) + \theta(-x^0)V(-x^0,\vec x)
\non\\
&=& V_+(X) + \epsilon(x^0)V_-(X)
\label{Vtilgen}
\ear
We will therefore give the final master formulas generalizing
eqs. (\ref{masterindirect}) and (\ref{Kdirmastermod}), (\ref{Kdirbulkmaster}), 
 (\ref{Kdirboundarymaster}) and list the short-time expansion of
the heat kernel in the half space.  In the following it should now be
understood that 
$D_i(p)\cdot D_j(p)= \sum_{\beta = 0}^{D-1} D^\beta_i(p)D^\beta_j(p)$, and
that only the zeroeth component of  $D_i(p)$ has a $p$ - part (when acting on $V_-$),
i.e. $D^\beta_i(p) = \partial^\beta_i + \delta^{\beta 0}ip_i$. Hence,
\bea
K^{\rm ind}_{\partial M}
&=& 
{1\over 4}
(4\pi T)^{-\frac{D-1}{2}}
\sum_{n=0}^{\infty}{(-T)^n\over n!}
\int_0^1d\tau_1 \cdots \int_0^1d\tau_n \int d^{D-1}x
\non\\
&&\hspace{-10pt}\times
\,\exp \Bigl[-{T\over 2} \sum_{i,j=1}^n \Bigl(G_{D,S}(\t_i,\t_j)
\vec{\partial_i}\cdot \vec{\partial_j} +G_{A}(\t_i,\t_j)
D^0_i(p) D^0_j(p)\Bigr)
\Bigr]\non\\
&&\hspace{-10pt}\times
\,
\prod_{k=1}^n 
\Bigl[V_+^{(k)}(0,\vec{x})+{1\over\pi i}\int_{\rm ev} {dp_k\over p_k}
V_-^{(k)}(0,\vec{x})\Bigr]
\label{masterindirectD}
\eea
is the master formula for the indirect term, whereas
\bear
K^{\rm dir}(T,X)
&=& 
(4\pi T)^{-{D\over 2}}
\sum_{n=0}^{\infty}
{(-T)^n\over n!}
\int_0^1d\tau_1 \cdots \int_0^1d\tau_n
\nonumber\\
&& 
\hspace{-10pt}\times
\,{\rm exp}\biggl[-{T\over 2}
\sum_{i,j=1}^n G_{D,S}(\tau_i,\tau_j)D_i(p)\cdot D_j(p)
\biggr]
\nonumber\\
&& 
\hspace{-10pt}\times
\,\prod_{k=1}^n
\, \Bigl\lbrack V^{(k)}_+ (X) + {1\over \pi i}\int_{\rm ev}{dp_k\over p_k}
\,\e^{ip_kx^0}V_-^{(k)}(X) \Bigr\rbrack
\label{KdirmastermodD}\\[4mm]
%\ear
%
%\bear
K_{M}^{\rm dir}(T,X)
&=& 
(4\pi T)^{-{D\over 2}}
\sum_{n=0}^{\infty}
{(-T)^n\over n!}
\int_0^1d\tau_1 \cdots \int_0^1d\tau_n
\nonumber\\
&& 
\hspace{-10pt}\times
{\rm exp}\biggl[-{T\over 2}
\sum_{i,j=1}^n G_{D,S}(\tau_i,\tau_j)\partial_i\cdot\partial_j
\biggr]
\prod_{i=1}^n
\, V^{(i)}(X)\,,\quad  (x^0>0)
\label{KdirbulkmasterD}\\[4mm]
%\ear
%
%\bear
K_{\partial M}^{\rm dir}(T)
&=& 
\half
(4\pi T)^{-{D\over 2}}
\sum_{n=0}^{\infty}
{(-T)^n\over n!}
\int_0^1d\tau_1 \cdots \int_0^1d\tau_n
\int_{-\infty}^{\infty}dx^0
\int d^{D-1}x
\nonumber\\
&& \hspace{-10pt}
\times \int_0^1dw \, \frac{\partial}{\partial w}
\,{\rm exp}\biggl[-{T\over 2}
\sum_{i,j=1}^n G_{D,S}(\tau_i,\tau_j)D_i(wp)\cdot D_j(wp)
\biggr]
\non\\
&&\hspace{-10pt}\times
\prod_{k=1}^n
\, \Bigl\lbrack V^{(k)}_+(X)+ {1\over \pi i}\int_{\rm ev}{dp_k\over p_k}
\,\e^{ip_kx^0}V_-^{(k)}(X) \Bigr\rbrack
\label{KdirboundarymasterD}
\ear
are respectively the master formulas for the complete local direct term, its
bulk part, and its integrated boundary part.

Finally we report the coefficients, up to order $9/2$, of the
short-time expansion for the heat-kernel trace, computed using the
above formulas with the SI Green's function
\bea
K_S(T) = (4\pi T)^{-\frac{D}{2}} \sum_{n\in {\mathbb N}/2} a_n T^n
\eea
with
\bea
a_0 &=&\int_M  1
\ccr[2mm]
a_{1/2} &=& 
\frac{\sqrt{\pi}}2\int_{\partial M} (\mp 1)
\ccr[2mm]
a_1 &=& \int_M (-V) 
\ccr[2mm]
a_{3/2} &=&
\frac{\sqrt{\pi}}2\int_{\partial M} (\pm V)
 \ccr[2mm]
a_2 &=& \int_M 
\frac{1}{2!}\, V^2 
+ \int_{\partial M} \frac{1}{2!}\, (\pm \partial_0 V)
\ccr[2mm]  
a_{5/2} &=&
\frac{\sqrt{\pi}}2\int_{\partial M} \frac{1}{2!}\, 
(\mp) 
 \left(V^2  -\frac{1}{2} \partial_0^2 V 
\right)\ccr[2mm]
a_3 &=& \int_M \frac{1}{3!}\,
\left( -V^3 -\frac{1}{2}(\partial_\beta V)^2 \right)
+ \int_{\partial M} \frac{1}{3!}\, 
(\mp) \left( 3 V \partial_0 V - \frac{1}{2}\partial_0^3 V \right)
\ccr[2mm] 
%%%%%%%%%%%%%%%%%%%%
a_{7/2}&=& \frac{\sqrt{\pi}}2\int_{\partial M} \frac{1}{3!}\,
 \Biggl[\mp\left(-V^3 
+\frac{3}{2} V\partial_0^2 V 
-\frac{1}{2} (\vec{\partial} V)^2
-\frac{3}{16}\partial_0^4 V \right)
+
\lx\{
\begin{array}{c}
-5\\ 7
\end{array}\rx\} 
\frac{3}{16}(\partial_0V)^2 \Biggr]
\ccr[2mm]
a_4 &=& \int_M \frac{1}{4!}\,
\left[ V^4+ 2 V(\partial_\beta V)^2 
+\frac{1}{5}(\partial_\beta\partial_\gamma V)^2\right]\non\\[1mm]
&&\hspace{-15pt}+
\int_{\partial M} \frac{1}{4!}\,
 \Biggl[\mp\Biggl( - 6 V^2\partial_0 V
+ 2V\partial_0^3 V
- 2 \vec{\partial}V\cdot\vec{\partial}\partial_0 V
 -\frac{1}{5}\partial^5_0 V \Biggr)
%\non\\[1mm]&&\hskip8cm
+
\lx\{
\begin{array}{c}
-9\\ 11\end{array}
\rx\}
\frac{2}{5} \partial_0 V \partial_0^2 V \Biggr]
\ccr[2mm]   
a_{9/2}&=& \frac{\sqrt{\pi}}2\int_{\partial M} \frac{1}{4!}\,
 \Biggl[ \mp\Biggl( V^4
- 3 V^2\partial_0^2 V
+ 2 V \big(\vec{\partial} V\big)^2  
+\frac{1}{5}\left({\partial_b}{\partial_c} V\right)^2
+\frac{3}{4}V\partial_0^4 V
\non\\[1mm]&&\hskip2cm
-\vec{\partial} V\cdot \vec{\partial}\partial_0^2 V
+\frac{5}{4} (\partial_0^2 V)^2
-\frac{1}{16} \partial_0^6 V 
\Biggr) 
+
\lx\{
\begin{array}{c}
5\\ -7\end{array}
 \rx\}
\frac{3}{4} V(\partial_0 V)^2
\non\\[1mm]&&\hskip2cm
+ \lx\{
\begin{array}{c}
7\\ -11 \end{array}
\rx\}
\frac{5}{64}\big(\vec{\partial}\partial_0 V\big)^2  +
\left\{
\begin{array}{c}
-7\\ 8 \end{array}
 \right\}
\frac{1}{4}\partial_0 V\partial^3_0 V
\Biggr]\non
\eea
where the upper coefficients in the braces refer to (space-time) Dirichlet boundary
conditions, whereas the lower ones refer to Neumann boundary
conditions. 

%\paragraph{Alternative form 1}

%%%%%%%%%%%%%%%%%%%%
%%%%%%%%%%%%%%%%%%%%
%%%%%%%%%%%%%%%%%%%%

%\vfill\eject
The same results, in the form produced by the DBC propagators, 
is obtained by adding to the $a_n$ suitable vanishing terms, written as  
total derivatives minus their boundary values. Here we list the first few ones
\bea
&& \Delta a_2=0=
-\int_M \frac{1}{3!} \Box V 
- \int_{\partial M} \frac{1}{3!} \partial_0 V 
\ccr
&& \Delta a_{5/2}=0= 
\pm \frac{\sqrt{\pi}}2\int_{\partial M} \frac{1}{3!} \vec{\partial}^2 V 
\ccr
&& \Delta a_3=0=
\int_M 
\frac{1}{3!} \partial_\beta \Big (V\partial_\beta V 
-\frac{1}{10} \partial_\beta \Box V \Big )
+
\int_{\partial M} \frac{1}{3!} \Big (V\partial_0 V 
-\frac{1}{10} \partial_0 \Box V \Big )
\ccr 
&& \qquad \qquad +
\int_{\partial M} 
\lx\{  \begin{array}{c}
8\\ -12
\end{array} \rx\} 
\frac{1}{5!}\partial_0 \vec{\partial}^2 V 
\ccr
&& \Delta a_{7/2}=0= 
\mp \frac{\sqrt{\pi}}2\int_{\partial M} \frac{1}{3!} 
\partial_b \left ( V\partial_b V - 
\Big ( \frac{1}{4} \partial_0^2 
+\frac{1}{10} {\vec{\partial}}^2 \Big ) \partial_b  V
\right )
\nonumber
\eea
that allow to obtain a consistent check with the coefficients computed
earlier in~\cite{Bastianelli:2006hq}.  
\end{appendix}
%%%%%%%%%%%%%%%%%%%%%%%%%%%%%%%%%%%%%%%%%%%%%%%%%%%%%%%%%
%%%%%%%%%%%%%%%%%%%%%%%%%%%%%%%%%%%%%%%%%%%%%%%%%%%%%%%%%

\vfill\eject

%\newpage

%%%%%%%%%%%%%%%%%%%%%%%%%%%%%%%%%%%%%%%%%%%%%%%%%%%%%%%%%%%%%%%%%%%%%%%%%%%%%%%


\begin{thebibliography}{99}

\bibitem{berkos}
Z. Bern and D.A. Kosower, \emph{Nucl. Phys. B} \textbf{379}, 451 (1992).

\bibitem{strassler}
M.J. Strassler, \emph{Nucl. Phys. B} \textbf{385}, 145 (1992).

\bibitem{ball}
R. D. Ball, \emph{Phys. Rept.} {\bf 182} (1989) 1.

\bibitem{vassilevich}
  D.~V.~Vassilevich,
  %``Heat kernel expansion: User's manual,''
  \emph{Phys. Rept.}  {\bf 388} (2003) 279
  [arXiv:hep-th/0306138].

\bibitem{kirsten}
  K.~Kirsten,
  \emph{Spectral functions in mathematics and physics,}
%\href{http://www.slac.stanford.edu/spires/find/hep/www?irn=4785843}{SPIRES entry}
Chapman \& Hall/CRC, 2001.  


\bibitem{review}
C. Schubert, \emph{Phys. Rept.} {\bf 355} (2001) 73, [arXiv:hep-th/0101036].

\bibitem{fhss}
D. Fliegner, P. Haberl, M.G. Schmidt and
C. Schubert, \emph{Ann. Phys. (N.Y.)} {\bf 264} (1998) 51 [arXiv:hep-th/9707189]. 

\bibitem{fss}
D. Fliegner, M.G. Schmidt and C. Schubert,
\emph{Z. Phys. C} {\bf 64} (1994) 111 [arXiv:hep-ph/9401221].

\bibitem{winnipeg}
D.~Fliegner, P.~Haberl, M.~G.~Schmidt and C.~Schubert,
  %``An improved heat kernel expansion from worldline path integrals,''
  \emph{Discourses Math. Appl.}  {\bf 4} (1995) 87
  [arXiv:hep-th/9411177].

\bibitem{qed}
M.G. Schmidt and C. Schubert, \emph{Phys. Lett. B} \textbf{318}
(1993) 438 [arXiv:hep-th/9309055]; 
D.G.C. McKeon and T.N. Sherry, \emph{Mod. Phys. Lett. A} \textbf{9} (1994) 2167;
D. Cangemi, E. D'Hoker and G. Dunne, \emph{Phys. Rev. D} \textbf{51}
(1995) 2513
[arXiv:hep-th/9409113];
V.P. Gusynin and I.A. Shovkovy, \emph{Can. J. Phys.} \textbf{74}
(1996) 282 [arXiv:hep-ph/9509383]; 
\emph{J. Math. Phys.} \textbf{40} (1999) 5406 [arXiv:hep-th/9804143]; 
R. Shaisultanov, \emph{Phys. Lett. B} \textbf{378} (1996) 354 [arXiv:hep-th/9512142];
M. Reuter, M.G. Schmidt and C. Schubert, \emph{Ann. Phys. (N.Y.)}
\textbf{259} (1997) 313 [arXiv:hep-th/9610191].

\bibitem{gaugetheory}
U. M\"uller, \emph{DESY}-96-154 [arXiv:hep-th/9701124].

%\cite{gra1}
\bibitem{gra1}
  F.~Bastianelli,
  %``The Path integral for a particle in curved spaces and Weyl anomalies,''
  \emph{Nucl. Phys.  B} {\bf 376} (1992) 113
  [arXiv:hep-th/9112035];
  %%CITATION = NUPHA,B376,113;%%
%\cite{Bastianelli:1992ct}
%\bibitem{Bastianelli:1992ct}
  F.~Bastianelli and P.~van Nieuwenhuizen,
  %``Trace Anomalies From Quantum Mechanics,''
  \emph{Nucl.\ Phys.\  B} {\bf 389} (1993) 53
  [arXiv:hep-th/9208059].
  %%CITATION = NUPHA,B389,53;%%

\bibitem{gra2}
F.A. Dilkes and D.G.C. McKeon, 
\emph{Phys. Rev. D} {\bf 53} (1996) 4388 [arXiv:hep-th/9509005];
 F.~Bastianelli and O.~Corradini,
  %``On mode regularization of the configuration space path integral in  curved
  %space,''
  \emph{Phys.\ Rev.\  D} {\bf 60} (1999) 044014
  [arXiv:hep-th/9810119];
  %%CITATION = PHRVA,D60,044014;%%
%\cite{Bastianelli:2000dw}
%\bibitem{Bastianelli:2000dw}
%  F.~Bastianelli and O.~Corradini,
  %``6D trace anomalies from quantum mechanical path integrals,''
  \emph{Phys.\ Rev.\  D} {\bf 63} (2001) 065005
  [arXiv:hep-th/0010118];
  %%CITATION = PHRVA,D63,065005;%%
  F.~Bastianelli and A.~Zirotti,
  %``Worldline formalism in a gravitational background,''
  \emph{Nucl.\ Phys.\  B} {\bf 642} (2002) 372
  [arXiv:hep-th/0205182];
  %%CITATION = NUPHA,B642,372;%%
%\cite{Bastianelli:2002qw}
%\bibitem{Bastianelli:2002qw}
  F.~Bastianelli, O.~Corradini and A.~Zirotti,
  %``Dimensional regularization for SUSY sigma models and the worldline
  %formalism,''
  \emph{Phys.\ Rev.\  D} {\bf 67} (2003) 104009
  [arXiv:hep-th/0211134];
  %%CITATION = PHRVA,D67,104009;%%
%\cite{Bastianelli:2005vk}
%\bibitem{Bastianelli:2005vk}
  F.~Bastianelli, P.~Benincasa and S.~Giombi,
  %``Worldline approach to vector and antisymmetric tensor fields,''
  \emph{JHEP} {\bf 0504} (2005) 010
  [arXiv:hep-th/0503155];
  %%CITATION = JHEPA,0504,010;%%
%\cite{Bastianelli:2005uy}
% \bibitem{Bastianelli:2005uy}
%  F.~Bastianelli, P.~Benincasa and S.~Giombi,
% ``Worldline approach to vector and antisymmetric tensor fields. II,''
  \emph{JHEP} {\bf 0510} (2005) 114
  [arXiv:hep-th/0510010].
  %%CITATION = JHEPA,0510,114;%%




\bibitem{bacozi}
F. Bastianelli, O. Corradini and A. Zirotti, 
\emph{JHEP} {\bf 0401} (2004) 023  [arXiv:hep-th/0312064].
  %%CITATION = JHEPA,0702,059;%%

%\cite{Bordag:2001qi}
\bibitem{Bordag:2001qi}
  M.~Bordag, U.~Mohideen and V.~M.~Mostepanenko,
  %``New developments in the Casimir effect,''
  Phys.\ Rept.\  {\bf 353} (2001) 1
  [arXiv:quant-ph/0106045].
  %%CITATION = PRPLC,353,1;%%

\bibitem{brane-casimir}
%\cite{Fabinger:2000jd}
%\bibitem{Fabinger:2000jd}
  M.~Fabinger and P.~Horava,
  %``Casimir effect between world-branes in heterotic M-theory,''
  \emph{Nucl.\ Phys.\  B} {\bf 580} (2000) 243
  [arXiv:hep-th/0002073];
  %%CITATION = NUPHA,B580,243;%%
%\cite{Garriga:2000jb}
%\bibitem{Garriga:2000jb}
  J.~Garriga, O.~Pujolas and T.~Tanaka,
  %``Radion effective potential in the brane-world,''
  \emph{Nucl.\ Phys.\  B} {\bf 605} (2001) 192
  [arXiv:hep-th/0004109];
  %%CITATION = NUPHA,B605,192;%%
%\cite{Nojiri:2000bz}
%\bibitem{Nojiri:2000bz}
  S.~Nojiri, S.~D.~Odintsov and S.~Zerbini,
  %``Bulk versus boundary (gravitational Casimir) effects in quantum  creation
  %of inflationary brane world universe,''
  \emph{Class.\ Quant.\ Grav.\ } {\bf 17} (2000) 4855
  [arXiv:hep-th/0006115];
  %%CITATION = CQGRD,17,4855;%%
%\cite{Hofmann:2000cj}
%\bibitem{Hofmann:2000cj}
  R.~Hofmann, P.~Kanti and M.~Pospelov,
  %``(De-)stabilization of an extra dimension due to a Casimir force,''
  \emph{Phys.\ Rev.\  D} {\bf 63}, 124020 (2001)
  [arXiv:hep-ph/0012213].
  %%CITATION = PHRVA,D63,124020;%%

\bibitem{wlmontecarlo}
H. Gies and K. Langfeld, \emph{Nucl. Phys. B} \textbf{613}, 353 (2001), [arXiv:hep-ph/0102185];
\emph{Int. J. Mod. Phys. A} \textbf{17}, 966 (2002), [arXiv:hep-ph/0112198].
\bibitem{wlcasimir}
H.~Gies, K.~Langfeld and L.~Moyaerts,
  %``Casimir effect on the worldline,''
  \emph{JHEP} {\bf 0306} (2003) 018
  [arXiv:hep-th/0303264];
  %%CITATION = HEP-TH 0303264;%%
H.~Gies and K.~Klingmuller,
  %``Worldline algorithms for Casimir configurations,''
  \emph{Phys.\ Rev.\ D} {\bf 74} (2006) 045002
  [arXiv:quant-ph/0605141].



\bibitem{HW}
  P.~Horava and E.~Witten,
  %``Heterotic and type I string dynamics from eleven dimensions,''
  \emph{Nucl.\ Phys.\  B} {\bf 460} (1996) 506
  [arXiv:hep-th/9510209];
  %%CITATION = NUPHA,B460,506;%%
  %``Eleven-Dimensional Supergravity on a Manifold with Boundary,''
  \emph{Nucl.\ Phys.\  B} {\bf 475} (1996) 94
  [arXiv:hep-th/9603142].
  %%CITATION = NUPHA,B475,94;%%

  

\bibitem{wlanomalies}
 L.~Alvarez-Gaume,
  %``Supersymmetry And The Atiyah-Singer Index Theorem,''
  \emph{Commun.\ Math.\ Phys.}  {\bf 90} (1983) 161;
  %%CITATION = CMPHA,90,161;%%
  L.~Alvarez-Gaume and E.~Witten,
  %``Gravitational Anomalies,''
  \emph{Nucl.\ Phys.\  B} {\bf 234} (1984) 269;
  %%CITATION = NUPHA,B234,269;%%
 D.~Friedan and P.~Windey,
  %``Supersymmetric Derivation Of The Atiyah-Singer Index And The Chiral
  %Anomaly,''
  \emph{Nucl.\ Phys.\  B} {\bf 235} (1984) 395.
  %%CITATION = NUPHA,B235,395;%%


\bibitem{McKean:1967}
H.~P. McKean and I.~M. Singer, 
%``{Curvature and the eigenvalues of the  Laplacian},'' 
\emph{J.\ Diff.\ Geom.} {\bf 1} (1967) 43.

%\cite{Kennedy:1979ar}
\bibitem{Kennedy:1979ar}
  G.~Kennedy, R.~Critchley and J.~S.~Dowker,
  %``Finite Temperature Field Theory With Boundaries: Stress Tensor And Surface
  %Action Renormalization,''
  \emph{Annals Phys.}  {\bf 125} (1980) 346.
  %%CITATION = APNYA,125,346;%%

\bibitem{Branson:1990a}
T.~P. Branson and P.~B. Gilkey, 
% ``{The asymptotics of the Laplacian on a  manifold with boundary},'' 
\emph{Commun.\ Part.\ Diff.\ Equat.} {\bf 15} (1990) 245.

%\cite{Cognola:1990kq}
\bibitem{Cognola:1990kq}
  G.~Cognola, L.~Vanzo and S.~Zerbini,
  %``A New Algorithm For Asymptotic Heat Kernel Expansion For Manifolds With
  %Boundary,''
  \emph{Phys.\ Lett.\ B} {\bf 241} (1990) 381.
  %%CITATION = PHLTA,B241,381;%%

%\cite{McAvity:1990we}
\bibitem{McAvity:1990we}
  D.~M.~McAvity and H.~Osborn,
  %``A Dewitt Expansion Of The Heat Kernel For Manifolds With A Boundary,''
  \emph{Class.\ Quant.\ Grav.}  {\bf 8} (1991) 603; (E) {\bf 9}
  (1992) 317; \emph{Class.\ Quant.\ Grav.}  {\bf 8} (1991) 1445; 
  %%CITATION = CQGRD,8,603;%%
%\cite{McAvity:1991xf}
%\bibitem{McAvity:1991xf}
%  D.~M.~McAvity and H.~Osborn,
  %``Asymptotic expansion of the heat kernel for generalized boundary
  %conditions,''
%  Class.\ Quant.\ Grav.\  {\bf 8} (1991) 
  %%CITATION = CQGRD,8,1445;%%
%\cite{McAvity:1992rf}
%\bibitem{McAvity:1992rf}
  D.~M.~McAvity,
  %``Heat kernel asymptotics for mixed boundary conditions,''
  \emph{Class.\ Quant.\ Grav.}  {\bf 9} (1992) 1983.
  %%CITATION = CQGRD,9,1983;%%

%\cite{Branson:1995cm}
\bibitem{Branson:1995cm}
  T.~P.~Branson, P.~B.~Gilkey and D.~V.~Vassilevich,
  %``The Asymptotics of the Laplacian on a manifold with boundary. 2,''
  \emph{Boll.\ Union.\ Mat.\ Ital.}  {\bf 11B} (1997) 39
  [arXiv:hep-th/9504029];
  %%CITATION = HEP-TH 9504029;%%
  K.~Kirsten,
  %``The a(5) heat kernel coefficient on a manifold with boundary,''
  \emph{Class.\ Quant.\ Grav.}  {\bf 15} (1998) L5
  [arXiv:hep-th/9708081].
  %%CITATION = CQGRD,15,L5;%%


%\cite{Branson:1999jz}
\bibitem{Branson:1999jz}
  J.~S.~Dowker and K.~Kirsten,
  %``Smeared heat-kernel coefficients on the ball and generalized cone,''
  \emph{J.\ Math.\ Phys.}  {\bf 42} (2001) 434
  [arXiv:hep-th/9803094];
  %%CITATION = JMAPA,42,434;%%
  T.~P.~Branson, P.~B.~Gilkey, K.~Kirsten and D.~V.~Vassilevich,
  %``Heat kernel asymptotics with mixed boundary conditions,''
  \emph{Nucl.\ Phys.\ B} {\bf 563} (1999) 603
  [arXiv:hep-th/9906144].
  %%CITATION = HEP-TH 9906144;%%

\bibitem{Bastianelli:2006hq}
  F.~Bastianelli, O.~Corradini and P.~A.~G.~Pisani,
  %``Worldline approach to quantum field theories on flat manifolds with
  %boundaries,''
  \emph{JHEP} {\bf 0702} (2007) 059
  [arXiv:hep-th/0612236].

%\cite{Bastianelli:2007jr}
\bibitem{Bastianelli:2007jr}
  F.~Bastianelli, O.~Corradini and P.~A.~G.~Pisani,
  %``Scalar Field with Robin Boundary Conditions in the Worldline Formalism,''
  \emph{J.\ Phys.\ A } {\bf 41} (2008) 164010
  [arXiv:0710.4026 [hep-th]].
  %%CITATION = JPAGB,A41,164010;%%






\end{thebibliography}
\end{document}